\documentclass[times]{acm-book}

\usepackage{amsmath}    
\usepackage{graphicx}   

\usepackage{hyperref}

\title{Advanced Topics for Information Retrieval}

\begin{document}


\mainmatter
\newcommand\blfootnote[1]{
    \begingroup
    \renewcommand\thefootnote{}\footnote{#1}
    \addtocounter{footnote}{-1}
    \endgroup
}

\setcounter{chapter}{13}

\chapter{Searching Personal Collections}

\chapterauthor{Michael Bendersky, Donald Metzler, Marc Najork, and Xuanhui Wang}

This chapter discusses search and discovery in personal collections. A personal collection is a set of digital assets associated with a particular individual (often referred to as the ``user'') or a limited set of individuals. Examples of digital assets include emails, notes, photos, files, bookmarks and the like. In the remainder of this chapter, we use the terms asset, item, and document interchangeably.

\blfootnote{This article appeared as Chapter 14 in Omar Alonzo and Ricardo Baeza-Yates (editors), Information Retrieval: Advanced Topics and Techniques, ACM Press, 2025, available at https://doi.org/10.1145/3674127.3674142}

During the first two decades of the personal computing age, these assets were typically stored on a computer controlled by the user or on a server controlled by the user's organization (say, a department's file server). With the advent of the Web as a computing platform, personal collections have increasingly moved from the devices controlled by individuals or their organizations into services operated by cloud service providers.  For email collections, such service providers include Outlook.com (originally Hotmail,  launched in 1996), Yahoo! Mail (originally RocketMail, launched in 1996), AOL Mail (originally NetMail, launched in 1997), Mail.ru (launched in 1998), and Gmail (launched in 2004). For file collections, cloud-based service providers include Box (founded in 2005), Dropbox (founded in 2007), Microsoft OneDrive (originally SkyDrive, launched in 2007), and Google Drive (launched in 2012). 

Personal collections differ from public collections (such as the universe of books, the set of all news stories, or the Web) in a number of significant ways:
\begin{itemize}
  \item Users are typically aware of most, if not all, of the items that exist in the collection. Thus, the typical retrieval task is refinding known items, as opposed to finding as-of-yet unknown documents that satisfy an information need.
  \item Users ``own'' the assets, and may spend effort on organizing them, in order to aid future refinding. Organization strategies include foldering (placing assets into a tree-based hierarchy), tagging (labeling assets with terms from an extensible vocabulary) and annotating (e.g. adding a textual description to a photo or video).
  \item Personal collections are typically many orders of magnitude smaller than public collections.  This shifts the trade-offs in a retrieval system. While most large-scale retrieval systems (e.g. web search engines) are optimized for precision, personal search tends to be optimized for recall---specifically, to ensure that when a user is searching for a known asset, that asset is among the retrieved assets. 
  \item User interactions are sparse. Personal corpora by definition contain assets that are personal to a user, and are typically not shared with other users.  This implies that the owner of an asset is the only person interacting with the asset, in contrast to user interactions with public corpora, where many users of a retrieval system may issue similar queries and interact with (for example, click on) the same result documents.  Web search engines typically leverage this interaction data both during the retrieval stage (e.g. by expanding a query with terms of related queries with co-clicked results, or by mining query reformulations across the user population) and the ranking stage (e.g. by using result clicks as a relevance signal for future similar queries).
\end{itemize}

These differences distinguish personal from public collections. The remainder of this chapter will examine the various facets in more detail.

\section{Organizing Digital Assets}
\label{sec:organizing-assets}

The advent of the personal computer enabled users to accumulate personal digital assets, and this technological development spawned research into the psychology of {\em personal information management}.  \citet{Lansdale1988Psychology} wrote a seminal perspective article on how psychological observations should influence the development of future personal information management systems. Lansdale studied the behavior of office workers and the techniques to manage paper-based information.  As observed earlier by~\citet{Malone1983How}, some office workers organize documents into well-structured collections (the ``filers'') while others simply stack them on their desks (the ``pilers'').  Contrary to first expectations, diligent filing of information does not always enable successful retrieval because hierarchical classification systems tend to have overlapping and thus somewhat ambiguous categories, making it difficult to locate the folder when attempting to locate a file.  As a result, workers may feel that there is an insufficient return on investment to filing new information, and thus simply add to unstructured piles. Lansdale concludes that a successful filing system  should be organized by attributes that users do recall and should support retrieval based on multiple attributes. This idea is borne out by tag-based filing systems, see Section~\ref{sec:label-based-org}.

\citet{Nardi+1994Filing} interviewed 15 users on their approaches to filing and finding information on their Macintosh computers, and the problems they encountered in the process. While they found that no two users in the study employed the same naming or foldering conventions, they also found that users tended to group files into three categories: ephemeral, archived and working. Ephemeral files have a short shelf life---their usefulness diminishes rapidly over time, and they will not be used thereafter. Working files are being actively used for a current task. Archived files are not actively used at the moment but have a long shelf-life, and a user may need to refind them months or years after they were created.  Nardi et al. observed that users tended to place ephemeral and working files into obvious and visible locations (e.g. the computer's desktop), while they relied on a combination of location-based retrieval (e.g. a well-structured folder hierarchy), attribute search and full-text search when refinding archived information.

\citet{Barreau1995Context} surveyed seven managers (who at the time of the study were the class of office workers most likely to receive and retrieve large numbers of files) on how they organized the information routed to them and retrieved it. Some of the documents curated by the users were stored locally on users' machines while others were stored on a remote file server. In addition to observing similar behaviors as \citet{Nardi+1994Filing} did, Barreau also found that users leveraged file ownership as an attribute in organizing and retrieving information. 

\citet{BarreauNardi1995Finding} synthesized the findings of the two previous studies, emphasizing their common observations: users prefered filing strategies that leverage spatio-visual cues (e.g. placing files on the desktop); users avoided elaborate filing schemes; users distinguished between ephemeral, working, and archived files; and finally, users accumulated relatively few archival files.

It is worth remembering that the above studies, conducted more than 25 years ago, examined the behaviors of the first generation of personal computer users.  Clearly there is a feedback loop between users and systems: information systems improve to better accommodate their user base, and users' behavior evolves in response.  

\citet{WhittakerSidner1996EmailOverload} study the strategies of 20 Lotus Notes users for filing and refinding email messages. Email, originally designed as a communication medium, has evolved into an archival medium as well as a means for task management, that is, people using email to assign work items to others or themselves, and subsequently using email to track task completion.  All three uses are asynchronous: replies to incoming emails may not be written immediately, email archival envisions a future need for the archived message, and task management anticipates future task completion. Consequently, users developed different strategies for managing this working information, all leveraging Lotus Notes folder mechanism. Many users found filing emails to be a cognitively challenging task, since creating folders and adding emails to folders requires users to anticipate future information needs.  Whittaker and Sidner introduce the notion of a {\em failed folder},  one whose name or purpose is too ambiguous to be useful or one too specific to effectively partition the information space.  The study identifies three fundamental filing approaches: some users avoid filing altogether and instead rely on search capabilities to refind needed information; some users are ``frequent filers'' whose inboxes tend to be short and contain mostly new messages; and some users are ``spring cleaners'' who periodically (e.g. every one to three months) clean up their inbox and archive items in folders. Importantly, the study found that ``spring cleaning'' is an ineffective strategy for information management, with over half of the folders of spring cleaners being failed folders.  

Two later studies by~\citet{Elsweiler+2008Memory,Elsweiler+2011Difficult} examined the impact of the filing strategies on how much users remember about individual messages. Somewhat surprisingly, they found that non-filers remember more details about their messages than spring cleaners do, who in turn remember more than frequent filers do.  

\citet{BoardmanSasse2004Stuff} performed a cross-tool and longitudinal study of personal information management strategies. Specifically, they studied the strategies of 31 participants for organizing their files, email messages, and web bookmarks.  The study confirmed observations of earlier studies (for example, users group into filers and pilers). Moreover, they observed that users are more invested into some classes of assets (such as their files and emails) than others (e.g. their bookmarks). Most interestingly, they observed that participation in the study caused introspection, resulting in two of the eight participants changing their filing behavior, both starting to separate working files from archival files.

\section{Labeling Digital Assets}
\label{sec:label-based-org}

As mention in Section~\ref{sec:organizing-assets} above, \citet{Lansdale1988Psychology} observed that it is often difficult to construct classification hierarchies that are non-ambiguous and where every item belongs to exactly one class.  The obvious alternative is to allow items to belong to many classes. In this model, assets are labeled with any number of {\em labels} or {\em tags}, and users can issue queries that retrieve all items labeled with a set of tags.  Labels are widely used in cloud-based email systems (e.g. Gmail labels), photo repositories (e.g. Flickr tags) and public messaging platforms (e.g. Twitter hashtags). But they are also available in most commercial file systems as a means of file organization, and there is a body of research on them.

\citet{Mogul1984Representing} proposed a file system where all file metadata is expressed as key/value pairs, akin to property lists in Lisp. The paper described how such a system could be implemented and what functionality the file system API would have to provide; it did not explore how this added functionality could enhance the recall of personal information.  

The Macintosh File System (MFS), released that same year, introduced the concept of bifurcated files: an MFS file consists of a data fork (containing its content) and a resource fork (containing its metadata). MFS evolved into the Hierarchical File System (HFS), launched the following year, which in turn evolved into HFS+ (launched 1998). HFS+ allowed for each file to have an unlimited number of forks.  Beginning with OS X 10.9, released in 2013, HFS+ forks allows users to attach an arbitrary number of custom tags to their files and to surface all files with a given tag through the Finder~\citep{Apple2013WhatsNew}.

The Windows NT File System (NTFS), released in 1993, supports custom properties.  In NTFS, a file is a collection of one or more data streams, each data stream consisting of a name (a key) and an associated byte stream (a value). In this model, the file content is the main data stream (keyed by {\tt \$DATA}), and each file property is an ``alternate data stream'' (ADS) with a name (the key) and a byte stream (the value). All file metadata is encoded using ADS. 

The Windows 7 operating system (released in 2007) exposed custom properties to users, allowing users to annotate files with custom properties (such as tags); however, the user interface for doing so was nonintuitive. To remedy this shortcoming, \citet{Cutrell+2006Flexible} introduced Phlat, a Windows shell for tagging, browsing, and searching digital assets (namely, files, emails and browsed web pages). Phlat used custom properties to store file tags, and Windows Desktop search to index and retrieve content.

Following Moguls' work, the idea of expressing file metadata as key/value pairs was further explored by~\citet{Gifford+1991Semantic}, who proposed the Semantic File System (SFS).  SFS employs a set of user-definable transducers to annotate each file with key/value attributes. Transducers are automatically invoked as files are being created or modified. In addition to making key/value pairs accessible through the file system API, Gifford et al. also introduced the concept of ``virtual directories'': Unix directories (folders) that contain all the files with a particular key/value pair combination. Again, the SFS paper did not explore how virtual directories could be leveraged for retrieval tasks.

The core idea of SFS was further refined by the Nebula file system~\citep{Bowman+1994FileSystem}.  All the metadata of a file (e.g. its name, aliases, type, and access control) are expressed as key/value pairs. Nebula introduces the concept of ``view'' (akin to SFS virtual directories) that groups together all files with a given key/value pair.  

There are a number of research papers that propose to use tags as a first-class mechanism for organizing file collections.  The TagFS file system~\citep{Schenk+2006TagFS,Bloehdorn+2006TagFS} was explicitly designed to explore tagging as the primary mechanism to organize a user's files. \citet{SeltzerMurphy2009hFAD} argue that users by and large do not remember where their files are stored, and they propose the {\em Hierarchical Filesystems are Dead} (hFAD), a filesystem namespace based on searchable tags.  

For organizing email messages, labeling was introduced by Pachyderm~\citep{Birrell+al1997Pachyderm} and became popular through the rise of Gmail, where labeling is the \emph{only} mechanism for users to organize their messages.  Photo repositories such as FlickR have used tagging both as a way to create a textual vocabulary for non-textual assets and as a way to make these assets discoverable and searchable.

\section{Automatic Classification and Clustering}

As mentioned in Section~\ref{sec:label-based-org}, the SFS~\citep{Gifford+1991Semantic} introduced the concept of transducers---heuristic analyzers that annotate files with labels based on their content.  Subsequent research advanced the idea of automatic classification of personal assets. 

\citet{Cohen1996Learning} described an application of his RIPPER rule learning algorithm to two email classification tasks: automatic detection of talk announcements (a binary classification task) and automatic foldering of email messages (a multinomial classification task). Cohen evaluated the system on three personal datasets and found that the rule-based approach outperformed a TF-IDF baseline.

In subsequent decades, there has been a large body of research on applying the modern machine learning technologies of the day to the task of personal content classification and clustering. For instance, \citet{Bekkerman:2004} conducted a benchmarking study comparing several popular text classifiers (of that time period), including Maximum Entropy, Naive Bayes, SVMs, and Winnow, for the task of automatic email foldering.

In a more recent example, \citet{Wendt+2016LabelProp} describe a multinomial classification algorithm for efficiently labeling machine-generated emails as belonging to a handful of categories (Social, Travel, Finance, etc.).  The system clusters machine-generated emails according to their generating template, leverages preexisting email classifiers to label some of these clusters with the majority label, then uses label propagation to propagate labels from majority-labeled clusters to topically or lexically similar unlabeled clusters, and finally uses an efficient lookup mechanism to assign labels to newly-arriving emails belonging to a known cluster.  These labels are surfaced in Gmail as ``Smart Labels.''

Office workers in particular tend to work on multiple work streams in parallel.  \citet{Kong+2020Learning} attempt to support this multithreaded activity by automatically grouping files into ``workspaces,'' each workspace representing the files related to a particular work stream.  Workspaces are orthogonal to the underlying filing system.  Clustering is based on item--item similarity, expressed in terms of lexical, metadata, and access pattern features.  The similarity function is not crafted by hand, but instead learned, using file co-access data as a weak supervision signal.  When the system infers a new workspace, it suggests it to the user and allows them to accept or reject it.

\section{Email Spam Filtering, Phishing Detection, Abuse Detection}
\label{sec:spam-phishing-abuse}

Email can be misused to transmit unwanted, abusive, or outright dangerous messages. There has been a great body of research on detecting such messages automatically for the purpose of discarding or flagging them.

According to Internet folklore, the very first email message was sent in 1971 by Ray Tomlinson to himself. Just four years later, \citet{Postel1975JunkMail} wrote a ``Request for Comment'' entitled ``On the Junk Mail Problem''; alas, this technical note did not concern itself with spam messages, but with denial-of-service attacks caused by misbehaving senders. \citet{Denning1982ElectronicJunk} anticipated the rise of junk emails and advocated for the research community to devise filtering mechanisms to stem the flow of unwanted messages.  However, as recounted by \citet{CranorLaMacchia1998Spam}, the first known ``email spam campaign'' did not take place until 1994. 

During these early years, a variety of techniques were proposed to combat junk emails, including using cryptographic techniques to create per-sender email addresses \citep{Gabber+1998Curbing} and thus effectively creating ``security by obscurity,'' by imposing nonnegligible computational~\citep{DworkNaor1993Combating} or monetary~\citep{Abadi+2003Bankable} costs on sending email messages.  However, none of these approaches found standards adoption; instead, email users rely on machine-learned classifiers to identify and filter out spam messages.

\citet{Sahami+1998Bayesian} were among the first to show that machine-learned classifiers are effective at filtering out spam messages.  In particular, they employed a Naive Bayes classifier and used lexical features (both terms as well as specific phases such as ``Free money''), punctuation features (e.g. copious use of exclamation marks), and metadata features (e.g. the domain of the sender).  \citet{Drucker+1999SVMforSpam} compared a number of different classification techniques and found that Support Vector Machines outperformed both boosted decision trees and learned classification rules---specifically, the RIPPER algorithm mentioned in the previous section.  \citet{Androutsopoulos+2000Experimental} compared spam filters based on manually curated ``black-listed'' terms to that of machine-learned classifiers, and they found that the Naive Bayesian classifier they considered far outperformed the keyword-based approach.  For a more comprehensive view of the state of ML-based spam classifiers, the interested reader is referred to a comprehensive survey by~\citet{Cormack2008EmailSpamFiltering} and a newer survey by~\citet{Dada+2019MLSpamFiltering}.

\citet{Spertus1997Smokey} considered the problem of hostile communications---messages that are inflammatory, hostile,or abusive in nature---and describes Smokey, a system for automatically filtering such messages from a user's email inbox. Spertus identified a number of signature features of hostile messages: sentences starting with ``You,'' use of imperatives, condescending language, profanities, insults, and epithets. These signatures are captured by a 47-element feature vector and supplied to a C4.5 decision tree-based binary classifier. 

A third class of undesirable emails are phishing messages---messages that count on the recipient trusting them, and abusing that trust for monetary or other gain. There are many flavors of phishing, ranging from generic messages that are sent to many millions of users to messages that are targeted at one particular user (so-called ``spear-phishing''). \citet{Fette+2007Phishing} describe PILFER, a binary classifier that predicts whether a given email is a phishing email or not.  PILFER uses content features of the email, including discrepancies between links and anchor text, whether any link contains a raw IP address, and whether some links go to ``non modal'' domains, as well as a number of features commonly used by spam classifiers. It also uses characteristics of any website linked from the email, such as the age of the website's domain registration. \citet{Nimeh+2007Comparison} compared the effectiveness of different classification techniques for phishing detection. They found that logistic regression provided the highest precision, while random forests performed best in terms of both recall and $F_1$ (the harmonic mean of performance and recall). The interested reader is referred to \citet{Hong2012StateOfPhishing} for a more in-depth survey of work on phishing email detection.

\section{Email Threading}
\label{sec:email-threading}

The previous two sections described tools and techniques to automatically process email messages---to file them, label them, or filter them.  All these techniques help reduce information overload.  Filtering out spam, phishing, and abuse also reduces annoyance, while filing and labeling make it easier to subsequently recover a message should the need arise.

Another way to impose structure on emails is by grouping the individual messages of a multiparty conversation into an {\em email thread}, and to surface conversations not at the level of individual messages but at the thread level. As we will discuss in Section~\ref{sec:refinding-using-search}, presenting messages in a threaded format is one of the most effective techniques for aiding refinding, and it has become the standard in all modern email experiences.  

Almost since the inception of email, RFC 680 \citep{MyerHenderson1975MessageTransmissionProtocol} has provided a dedicated header (IN-REPLY-TO) for specifying the tree-structure of a conversation; however, the standard did not specify what the legal values of this header are, and therefore its implementation was fragmented. This situation was widely understood and well-described for instance by \citet{Lewis+1997Threading}, who then proposed to use information retrieval techniques (namely, lexical similarity of messages in the same thread due to quoted texts) to provide a work-around. Parallel attempts up to and including RFC 2822 \citep{Resnick2001InternetMessageFormat} did not manage to resolve this ambiguity. As a result, implementors of email clients had to reverse-engineer the various competing implementations of the IN-REPLY-TO header and resolve ambiguities in the client.  \citet{Zawinski2002MessageThreading} provides an excellent account ``from the trenches'' of what was involved in achieving this, and a specific heuristic for performing this task. Zawinski's heuristic for reconstructing threads has been adopted by RFC 5256~\citep{CrispinMurchison2008SortAndThread}.  In spite of this, recovering thread structure continued to be a subject of research; for example, \citet{YehHarnly2006Reassembly} describe how to use undocumented Microsoft Exchange header information to recover threads, and in the absence of such information, how to exploit similarity between quoting and quoted portions of a message thread in assembling the thread.

All of the above is concerned with grouping messages into a forest according to their in-reply-to relationship, such that a message is a child of another message if it was a (direct) response to the parent message.  \citet{Ailon+2013Threading} refer to this as {\em syntactic threading}, and they propose the higher-level notion of {\em causal threads}: chains of email conversations that are causally related but not part of the same syntactic thread.  For example, consider the scenario of purchasing an item from Amazon. Amazon may send you an email confirmation upon order, a second email once the item has shipped, and a third email upon delivery. Syntactic threading will not aggregate these three emails into the same thread. However, this {\em ordered--shipped--delivered} chain of emails is a common pattern and can be learned. Ailon et al. do so by performing template induction over the Yahoo email corpus, annotating machine-generated emails with their template IDs, mining the annotated corpus for temporal patterns (e.g., ``users are likely to receive a shipment confirmation shortly after an order confirmation''), and using these discovered relationships to group conversations into causal threads.

\section{Known-item Retrieval}
\label{sec:known-item-retrieval}

Early information retrieval research was heavily informed by publication search, and it was the Library Science community that coined the term {\em known-item search}.  \citet{Swanson1972Requirements} defined known-item search as the attempt to locate a specific document that is known to exist.  He distinguishes known-item search from {\em subject search}, the attempt to locate all available information on a given subject or topic, and {\em reference search}, the attempt to find the answer to a specific (often factoid) question.  Swanson's article draws on earlier ethnographic studies, for example, an M.A. thesis by~\citet{Vaughan1968Effectiveness} that found that gaps in human memory recall (e.g. the precise title of a book) make known-item search a challenging problem.

\citet{Allen1989Recall} focused on the role of recall cues in known-item retrieval by examining the ability of 44 subjects to recall bibliographic, structural, and content cues of a short research paper they read three weeks earlier. Perhaps not surprisingly, the subjects performed worst on recalling bibliographic cues of the research paper. This finding is the ``raison d'etre'' for known-item retrieval systems: users are much more likely to recall the contents of an asset than its formal descriptors---its name, filing location, or the like.

\citet{Elsweiler+2007Towards} conducted a diary study to support the hypothesis that the root cause of unsuccessful known-item searches is a lapse in memory. Twenty-five participants recorded memory lapses over the course of one week, capturing 261 lapses. The majority of these memory lapses were retrospective---the failure to recall a previously memorized bit of information. Subjects employed a variety of preventative and recovery strategies to cope with such lapses. Elsweiler et al. translated these findings into a set of principles to improve the design of Personal Information Management tools: multimodal search capabilities, support for ``mental retrieval journeys,'' feedback cues about past interactions with surfaced results, and annotation facilities that allow users to encode newly remembered information in line with retrieval.  Some of these ideas are further expanded in Section~\ref{sec:episodic-memory}.

\section{Refinding Using Search}
\label{sec:refinding-using-search}

Full-text search is one of the most powerful tools for locating items in a collection.  As mentioned in Section~\ref{sec:known-item-retrieval}, the dominant search task for personal collections is known-item search: refinding an item that is known to be a part of the collection, be it a file, an email, or a visit to a website. Consequently, researchers have studied how to support known-item search in the context of personal search.

To gain a better understanding of such refinding tasks, \citet{ElsweilerRuthven2007Towards} asked 36 participants to record details of all email and web page refinding tasks they embarked on over a period of three weeks. They then labeled each search task according to whether the participant is looking for (1) a single nugget of information contained in a document that may or may not be known (lookup task), (2) a single document that is known to the user (item task), or (3) multiple documents that collectively answer the information need (multi-item task). For email searches, they found that the majority are lookup tasks, and lookup and item tasks combined account for about 95\% of all tasks. For web search, about 53\% were observed to be item tasks, and lookup and item tasks accounted for about 86\% of all web search tasks.

\citet{Dumais+2003Stuff} describe ``Stuff I've Seen'' (SIS), a system that records and indexes the information that a user is exposed to, including personal assets such as files and emails as well as personal exposure to public assets (e.g. web page visitations). SIS was evaluated through a user study that involved 234 participants who used the system for 47 days. To obtain qualitative insights, participants completed a questionnaire prior to using the system and again toward the end of the study; to obtain quantitative insights, interaction logs were collected throughout the study. One of the most surprising findings of the study was that participants preferred search results to be ranked by date rather than by relevance (SIS used lexical features and a BM25-based ranker). The authors speculate that users have a rough idea when they received, modified, or interacted with the item they are looking for, and they can therefore leverage dates effectively when skimming through the result list.  This observation will be further explored in Section~\ref{sec:episodic-memory} below.

\citet{Whittaker+al2011Wasting} studied users' email refinding strategies in BlueMail~\citep{Tang+CHI2008TagIt}, an email client that supports filing, tagging, threading, and full-text search. Filing and tagging emails incur an up-front cost as email is received, while using full-text search and leveraging message threading do not. The authors found that opportunistic search behaviors that do not rely on foldering and tagging dominated refinding activities. They also found that refinding sessions that involved search and those that leveraged conversation threading were more successful than sessions that did not. 

More recently, \citet{Ai+2017Characterizing} performed a query-log analysis of searches issued to the Microsoft Outlook cloud-based email service.  The log contained about 2 million queries issued by 711,000 users over a span of one week.  Most of the activities came from desktop clients of enterprise users. The authors also surveyed 324 of these users to obtain qualitative insights. The study confirmed that users' search tasks were overwhelmingly refinding of known items.  Moreover, the study found that compared to web search, queries were shorter and fewer results were clicked. Finally, repeated queries were common, but different results were clicked in different sessions. 

Large-scale studies along the lines of \citet{Ai+2017Characterizing} are difficult to conduct due to user privacy concerns, which make it all but impossible to analyze consumer email search behavior. \citet{Mackenzie+2019Exploring} proposed a novel approach to overcome this problem: rather than analyzing real consumer search behavior, they generate simulated searches by assigning email refinding tasks to workers on a crowd-sourcing platform, leveraging the workers' own email corpora. The worker is first presented with a target email drawn from their own email collection and then asked to refind it using a search interface. The authors found that workers employed a number of different query reformulation strategies to locate the target item, including typo correction, partial and full rewrites of the query, specialization and generalization, and occasional reversion to an earlier formulation. Query reformulations were common (with an average of about two reformulations per refinding task), suggesting that personal search indeed suffers from recall issues. The authors also examined the impact of different result orderings: time-based, relevance-based, and hybrid. Not surprisingly, they found that time-based email ordering led to more unsuccessful searches as the age of the target email increases. 

Motivated by the recall issues in personal search, there were several studies on how to augment user queries to improve recall \citep{Kuzi+al:2017,Li+2019MultiView}. For instance, \citet{Li+2019MultiView} developed a method to find domain-specific synonyms for query terms that leverages user activities (query reformulations and query--result interactions) as well as word embeddings to find semantically related query terms. They validated their query expansion approach on Gmail search traffic and found that expanded queries experienced higher success rates, as measured by result click-through rates and click positions.

\section{Leveraging Episodic Memory}
\label{sec:episodic-memory}

As discussed in Section~\ref{sec:known-item-retrieval} above, the inability of users to locate an asset known to be in their collection can be viewed as a memory failure.  It is not uncommon for users to forget the {\em location} of an asset (or key terms in a document that may be used to search for it) but instead to recall ancillary information that is not useful for retrieval, such as events that co-occurred when the user last interacted with the asset. The ability to recall events in time is referred to as {\em episodic memory}.  Naturally, researchers have tried to leverage episodic memory for retrieval tasks.

\citet{LansdaleEdmonds1992Using} describe MEMOIRS (``Managing Episodic Memory for Office Retrieval Systems''), a prototype retrieval system designed to exploit event memory.  MEMOIRS logs all interactions between users and their digital assets, thereby creating a timeline (a diary) of ``digital life events.'' Recording is automatic, and the timeline is immutable---events cannot be edited or deleted by the user. Users can browse the timeline and see all their asset interactions during a specific time window, and they can restrict the types of assets and interactions that are shown. 

The idea of using timelines as a way to organize and retrieve digital assets was further explored by the Lifestreams system~\citep{FreemanFertig1995Lifestreams,FreemanGelernter1996Lifestreams}.  While a MEMOIRS diary is a time-ordered sequence of interaction events with assets, a Lifestream is a time-ordered sequence of the assets themselves.  Assets (files, emails, calendar entries, etc.) are added to a user's Lifestream when they are received.  Lifestream supports user-definable agents that are triggered by events (e.g. by an interaction with an asset), and the invoked agent may place a clone of the document into the Lifestream (thereby emulating the functionality of MEMOIRS). Lifestreams can be browsed, filtered, and searched through an index.

\citet{Ringel+2003Landmarks} explore how episodic memory can be leveraged in a search system.  Specifically, they studied how the search results of an existing personal information retrieval system (SIS, described in Section~\ref{sec:refinding-using-search} above) can be augmented with ``landmark events'' that anchor each result in time, and to what extent such landmarks support retrieval tasks. Like MEMOIRS, their system captures a user's interactions with their digital assets (i.e. personal events); the system also maintains a timeline of public events (holidays, news stories).  Results to a search query are displayed on a vertical timeline, and adjacent to them the system displays landmark events, both public (e.g. a news event) and personal (e.g. a photo that was taken around that time).  The authors found that adding landmarks to the chronological display of search results improved retrieval; both quantitatively (measured by task completion time) and qualitatively (measured by surveying ten participants of a user study).

Below, in Section~\ref{sec:digital-memory} on ``Human Digital Memory,'' we will further explore the idea of capturing the entirety of a user's interactions with the world and providing tools to retrieve these interactions.

\section{Test Collections}
\label{sec:test-collections}

The Information Retrieval research community has long relied on shared test collections as a way to perform experiments that can be repeated and be reproduced by other research groups, and that allows researchers to compete on a common research challenge and measure progress over time. Unfortunately, while there are a number of test collections based on public documents, there are but a few collections for personal retrieval tasks.

One such collection is the Enron corpus \citep{KlimtYangCEAS2004Introducing,KlimtYangECML2004Enron}, a set of 619,446 messages belonging to 158 individuals, all of them managers and executives of Enron. The email messages became available as part of the legal discovery process during Enron's bankruptcy proceedings.  However, while many research papers have used this corpus, it is surrounded by some ethical concerns, chiefly the fact that the individuals whose messages are included did not provide their informed consent.  Also, note that the Enron data set contains emails, but neither queries nor relevance judgments.

A second test set for email consists of a collection of 198,394 messages sent to a number of mailing lists of the World Wide Web Consortium (W3C), and these messages were recovered by the organizers of the Text Retrieval Conference (TREC) to provide a corpus for the email search task of the TREC-2005 Enterprise track \citep{Craswell+2005Overview}. Along with the collection comes a set of 150 queries (framed by the TREC organizers, not issued in a natural context) along with the one item in the collection that is deemed the correct result. Unfortunately, at the time of this writing, this collection is no longer publicly available.

The Avocado research email collection \citep{Oard+2015Avocado} is the newest and largest such corpus. To quote the researcher who provided the collection, it ``consists of emails and attachments taken from 279 accounts of a defunct information technology company referred to as {\em Avocado}.'' The corpus contains about 2 million items. 
Separately, members of the same research group constructed a test set of 65 queries, each with about 100 results obtained by pooling the top 25 results of 18 retrieval ``systems'' (actually all implemented in Terrier), and each result labeled with its relevance and sensitivity judgments \citep{Spertus1997Smokey,Sayed+SIGIR2020TestCollection}.

\section{Crafted Rankers}
\label{sec:crafted-rankers}

As mentioned in Section~\ref{sec:refinding-using-search}, ~\citet{Dumais+2003Stuff} found that users of the SIS system preferred chronological ordering of search results over relevance-based ordering, lending support to the argument that episodic memory aids in retrieval tasks.  On the other hand, \citet{Mackenzie+2019Exploring} found that when presenting results in chronological order, the rate of unsuccessful searches was correlated to the age of the target item, arguing that there is still a need for relevance-based ranking.

Scoring functions such as the cosine similarity of TF-IDF vectors, BM25F and PL2F are examples of {\em crafted} rankers, constructed by Information Retrieval researchers based on underlying models of relevance (e.g. the term vector-space model or the probabilistic relevance model). The BM25 ranking model proposed by~\citet{RobertsonWalker1994SomeSimple} has been the workhorse of lexical ranking models for a long time. \citet{Robertson+CIKM2004Simple} extended BM25 to ``fielded'' documents---structured documents where some portions (such as their title or section headers) are more significant than others. \citet{MacdonaldOunis2006Combining} propose to use a similar fielded model PL2F for email search.  They evaluate this collection on the known-item search task of the TREC 2005 Enterprise track, which is based on a corpus of (public) messages sent to W3C mailing lists.

File retrieval systems (whether desktop-based or cloud-based) operate over a multitude of different file types, such as PDF documents, Word and Excel files, or email messages stored on a user's machine.  These file types differ in their structure and their associated metadata. Fielded relevance models such as BM25F and PL2F are tuned to assign specific weights to each field, and these models may not perform optimally when used on a heterogeneous collection. \citet{KimCroft2010Ranking} investigated this issue. They propose to use type-specific probabilistic retrieval models for each file type, resulting in one result per type, and to merge these result lists into a final list using the CORI collection ranking metric~\citep{Callan+1995CORI}.

\citet{AbdelRahman2010NewApproach} describes an email retrieval system that leverages the threaded nature of email conversations.  Prior to indexing, the system groups messages into threads using the heuristic by \citet{YehHarnly2006Reassembly} described in Section~\ref{sec:email-threading}. Conversations are then indexed using an algorithm due to~\citet{Broder+2006Indexing} (more fully described in Section~\ref{sec:infrastructure}), which maps conversations to ``document trees'' to capture the repeating nature of quoted previous messages in a conversation. The ranking function combines content, subject, and sender scores. The content score is the TF-IDF cosine similarity between query and message, the subject score is a binary match between query and subject line, and the sender score is the TF-IDF cosine similarity between the query and all messages ever received from that sender.

\section{Associative Browsing and Searching}
\label{sec:associative-browsing}

Modern search and recommender systems rely on a multitude of signals when retrieving and ranking search results or recommending items. These signals include the lexical similarity between query (or context) and item, linkage between items (e.g. hyperlinks between web pages), and user--item interactions. But while such signals are abundantly available to (say) a major web search engine, they are less so for a personal retrieval system.  In particular, many types of personal collections lack explicit linkage between items, and interaction signals will be sparse due to the personal nature of the collection, where an item can only be accessed by its owner.  This section looks at attempts to overcome this situation.

\citet{SoulesGanger2005Connections} describe Connections, a file system search tool that infers linkage between documents based on temporal co-access patterns and that uses link-based in addition to lexical features when retrieving and ranking search results. The system maintains a ``relationship graph'' where each node represents a file and each weighted edge represents the number of times two files have been co-accessed within (say) a half-minute window. The relationship graph is leveraged both during retrieval and ranking. During retrieval, an initial result set is obtained using standard lexical similarity between query and items, and then expanded by adding for each result all items that are nearby in the relationship graph, using edge weights and path lengths to determine proximity. The ranking phase uses the Indri ranker to perform an initial ranking, and then uses link-based signals (including PageRank and HITS) computed over the relationship graph to re-rank the results.  The system was evaluated on a small test set (the results to 25 queries, assessed by six judges), and it outperformed Indri as a baseline system.

\citet{DongHalevy2005Platform} describe SemEx (``Semantic Explorer''), another personal retrieval system that creates associations between items and leverages these associations during retrieval.  Unlike Connections, which inferred associations based on file co-access patterns and has but a single type of association, SemEx has a rich scheme of association types and uses handcrafted rules to extract them. Nodes in the SemEx graph can be items (emails, documents, etc.) as well as entities contained in the items (e.g. a person's name or email address); and edges are labeled with predicates (e.g. {\sl authorOf}). In effect, SemEx attempts to extract a Personal Knowledge Graph from the items in a user's collections. This rich graph allows for ``browsing by association,'' a powerful new paradigm for exploring one's personal assets, and also allows for more powerful queries.  The downside is that SemEx assumes the existence of powerful association extraction tools---crafting such tools is a laborious and brittle process (e.g. metadata formats of specific file types may change over time).

\citet{Chirita+2006Beagle++} describe Beagle++, a desktop search tool that combines ideas from Connections and SemEx. It builds on top of Beagle, the desktop search engine that is part of the Gnome family of desktop tools for Linux. Beagle++ aims to improve Beagle's Lucene ranker by incorporating semantic information about the files in the corpus.  Like Connections, it monitors interactions between users and items, and like SemEx, it extracts metadata such as author name from documents.  This information is stored as RDF triplets, forming an RDF graph. Beagle++ computes PageRank-inspired ``ObjectRank'' scores for each item in the graph. At query time, ObjectRank scores are used to reweigh the TF-IDF scores returned by the Lucene ranker.

\citet{Chen+2009SearchYourMemory} describe XSearcher, later renamed to iMecho \citep{Chen+2009iMecho}, another desktop search engine that attempts to infer associations between items in the file system. iMecho uses multiple signals to infer associations: (1) lexical similarity between items, (2) items being co-located in the same folder, (3) explicit user activities (e.g. a user copying a file), and (4) implicit user activities such as co-access patterns. iMecho computes ObjectRank scores over the resulting graph and uses these scores at retrieval time to re-rank results.  Unlike Beagle++, users can adjust the strength of each association type at query time, resulting in customizable rankings.

\citet{Kim+2010Building} describe LiFiDeA, another desktop search system that automatically creates associations between items and allows the user to leverage them for browsing and searching.  LiFiDeA has two classes of items: actual documents (e.g. email, files and web sites) and {\em concepts} (e.g. authors, places or key terms). Some concepts such as the sender of an email are automatically extracted from documents while others are provided by the user. For example, a user can promote a query (which retrieves a set of documents), a tag attached to a document, or a document itself (e.g. a Wikipedia page) to be a concept. Concepts are linked to the documents they are extracted from or associated with.  The LiFiDeA user interface allows browsing of both document and concept space, and to pivot from one to the other. In follow-on user studies, \citet{Kim+2011Evaluating} found that this associate browsing model indeed promotes known-item finding tasks.

\section{Learning to Rank for Personal Collections}
\label{sec:learning-to-rank}

An alternative to crafting rankers based on theories of document relevance is to automatically {\em learn} rankers from examples. In this context, an example consists of a query and the corresponding results set, each result having a binary or graded relevance label. Learning-to-Rank, first proposed by \citet{Freund+ICML1998Combining} and brought to the fore by \citet{Burges+ICML2005LTRusingGD}, is a vibrant area of research within the Information Retrieval community; the interested reader is referred to \citet{Liu2011LTRforIR} for a comprehensive overview of the early years.

\citet{Aberdeen+2010Learning} describe the learning algorithm behind Gmail Priority Inbox, a feature that displays the most {\em important} messages broken out at the top of the message list.  It is worth emphasizing that this is not a relevance problem since priority messages are shown absent of any query or any explicitly stated information need. Priority Inbox uses past user interactions as a weak supervision signal for model training. The model is based on simple logistic regression, and each message is represented by content features (i.e. textual features of subject and body), thread features (did the user interact with this thread before?), and label features (was the message auto-labeled using a user-specified rule?). Since importance is a highly subjective notion, Priority Inbox trains a global model over the entire user population and a personalized additive model for each individual user.  It also automatically tunes the per-user threshold above which a message is shown, based on the user's past interactions.

\citet{Carmel+2015Rank} revisited the question of whether email users preferred chronological or relevance-ordered search results. The authors present ``Relevance eXtended'' (REX), a learned ranking function for the email domain. REX uses BM25F to estimate the relevance of an email to a query, and furthermore it uses a number of query-independent importance signals: message (message freshness, user interactions with the message, presence of attachments, etc.), sender features (e.g. do this user and others actively correspond with the sender?), and recipient features (e.g. was this message sent to the user alone or to a group?).  These features are combined using SVMRank~\citep{Cao+2006Adapting}, trained over a set of queries sampled from search logs. Each training example consists of two result messages returned for the same search, one clicked and the other not.  In other words, REX uses clicks as a weak supervision signal.  An editorial evaluation showed that users preferred REX over chronological ordering. 

The approach taken by Carmel et al. of using result clicks as weak supervision signals ignores the fact that users exhibit a strong bias toward results near the top of the result list, and are far less likely to click results further down, even if these results are relevant.  It is crucial to correct this bias when using interaction data as relevance labels. \citet{Wang+SIGIR2016LTRwithBias} describe a technique for debiasing such log-based training examples.  The basic idea is to quantify the bias of each position (whether globally or query-dependent) by exposing a sample of users for a short period of time to randomized results. With users' bias for each position in the result list now known, the training data can be debiased using inverse propensity scoring.  \citet{Joachims+WSDM2017Unbiased} explored this idea further, and in the process devised a less disruptive procedure for quantifying users' bias for each rank: rather than fully randomizing the result lists for some sample of users and some period of time (and thereby severely degrading their experience), just swap the top document with another randomly selected document in the list (a much less disruptive intervention). Subsequent works devised multiple techniques for estimating the position bias directly from biased click logs, without the need for exposing a sample of users for some period of time to degraded ranking experiences \citep{Wang+WSDM2018PBEstimation,Ai+SIGIR2018UnbiasedLTR,Agarwal+WSDM2019EstimatingPB}.

All of the ranking models described in the previous paragraphs were based on LambdaMART~\citep{Wu+2010BoostingforIR} (a tree-based model) or some flavor of Support Vector Machines (a linear model).  In recent years, there has been a great amount of interest in using deep neural networks for ranking tasks, but until recently they have been unable to replace LambdaMART, and merely served as second-stage additive scorers. \citet{Li+2019Combining} provide an example of such stacked rankers, evaluated on email and file search tasks.


\section{Desktop Search}
\label{sec:desktop-search}

Full-text search is one of the most powerful means for locating information, but it was applied to file systems relatively recently. The earliest work we are aware of is Glimpse \citep{ManberWu1994Glimpse}, a system for indexing and searching Unix file systems.  Glimpse indexes individual words, but instead of a standard inverted index it uses a two-level indexing scheme that leads to smaller indices at the expense of slower queries. 

Today's mainstream PC operating systems provide integrated full-text search over the local file system.  Microsoft provided Indexing Service as early as 1996 and Windows Desktop Search in 2003. Apple introduced the OS X Spotlight search engine in April 2005. All of these search systems provide efficient retrieval of files containing the query terms; however, they do not provide any mechanisms for increasing recall (e.g. by expanding queries with synonymous terms) nor do they provide and form of relevance-based ranking.

\citet{Aharony2007Ranking} and \citet{Cohen+2008Ranking} studied three approaches to ranking desktop search results: a baseline crafted ranker (see Section~\ref{sec:crafted-rankers}) that uses content (via TF-IDF) and metadata features (access file, file path, file size, etc.); three learned ranking functions (see Section~\ref{sec:learning-to-rank}),  the best being SVM-based; and another crafted ranker leveraging insights from the learned models. Interestingly, the learning rankers were trained on individual users’ logs.

There has been some research on creating reusable test collections for personal search tasks. \citet{Chernov+2007Building,Chernov+2008Evaluating} proposed a pooling-based technique for assembling a test collection for desktop search, and the privacy considerations that go along with such an endeavor. The authors did indeed create such a collection and used it for subsequent research, but as far as we know did not make it available to the research community. \citet{KimCroft2009Retrieval} proposed a different technique for creating such a reusable test collection for personal search tasks. Starting from the assumption that personal files are people-centric, the authors reuse a set of names contained in publicly accessible W3C discussion groups, then used a commercial search engine to collect documents for each named individual, and finally synthesized known-item queries targeting this collection.

\section{File Recommendation}
\label{sec:file-recommendation}

Search and recommendation are closely related.  Both types of systems present a list of items to a user. In search settings these items are relevant with respect to an explicit query; in recommendation settings they are relevant given the user's context (e.g. previous consumption patterns).  It is therefore natural to consider file recommendations.

\citet{Tata+2017QuickAccess} describe Quick Access, a service in Google Drive (a major cloud-based file system) that recommends to users the items they may need to access next.  The recommender system is based on a simple feed-forward deep neural network.  Each item is represented by a fine-grained time-series of past interaction events (e.g. open, edit, and comment) on that item, as well as other contextual information such as upcoming calendar events, number of items in a user's collection, and whether the account is a private or work account.  The authors report an 18.5\% increase in recommendation accuracy over baseline heuristic (recommending the $k$ most-recently-used items) for highly active users, and they further report that on average users spend 50\% less time locating the file they wish to open.  In a follow-up paper, \citet{Tata+2019ItemSuggest} describe the architecture of ItemSuggest, the recommender engine underlying the Quick Access service. The authors pay particular attention to how ItemSuggest manages data integration from many disparate sources (e.g. Drive activity logs and Calendar state) and how it supports the rapid exploration of new features and new model architectures. In a third paper, \citet{Chen+2020Improving} describe model quality improvement work (better candidate generation, improved DNN architecture, and latency improvements) they performed over a period of multiple years, leading to another 10.7\% increase in accuracy.

In order to better understand how users interact with file recommender systems, \citet{Xu+WWW2020Understanding} studied large-scale logs of user interactions with the Microsoft Office.com Recommended Document Pane (RDP). RDP is similar to Quick Access and moreover provides a short explanation of why an item was recommended. The 14 predefined explanations can be broken down into collaborative explanations (e.g. the item was shared by a collaborator) and individual explanations (e.g. you recently opened the item).  The study measures user interactions in terms of click-through rate (CTR) and recognize rate (RR). While CTR is a raw interaction measure, RR measures whether the user recognized a useful item as it was recommended. The study found that there is a strong correlation between a recommendation's explanation and its recognition rate. Among collaborative explanations, {\sl CommentByOthers} had the highest RR; among individual explanations it was {\sl EditByYou}.
Another interesting finding of the study was that since recommendations are arranged in a linear list, they suffer from position bias similar to what has been observed for search results (see Section~\ref{sec:learning-to-rank}). Furthermore, once users have engaged with the system, their engagement remains high for the duration of the session. \citet{Jahanbaksh+CHIIR2020Effects} followed up on this large-scale quantitative study by performing a smaller-scale qualitative study, surveying 108 participants.  The study provided deeper insights into the factors that drive successful interactions with file recommendations.  Most noteworthy, the study found that presentation only influenced recognition time (a finding that is at odds with \citet{Tata+2017QuickAccess} who found that it drove engagement) and that users are more likely to recognize items they interacted with either recently or intensely. However, users professed that they were more ``interested'' in documents that were older and that they had interacted with less.

\section{Infrastructure for Cloud-based Personal Collections}
\label{sec:infrastructure}

Initially, personal collections tended to physically reside on a computer controlled by the owner of the collection.  With the rise of cloud-based email and file storage services, this is no longer the case. Migrating collections to cloud-based services has many advantages---greater data stability (users do not have to worry about backups), accessibility from multiple devices, economies of scale, and ``wisdom of crowds'' effects resulting from ML models trained in a privacy-preserving fashion on the data and activities of large user populations.  But these added capabilities also entail extra complexities: the dependence on centralized services, the reliance on network connections, the need for proper access control, and the dangers of privacy leaks. Unfortunately, not much has been published on systems infrastructure for cloud-based personal collections.

When indexing the collections of many users, a central service could maintain a separate inverted index per user (so access control can be handled at the collection level), or it could pool the documents of all users into a single inverted index (so access control has to be performed at the document level). The interested reader is referred to \citet{Hawking2011EnterpriseSearch} for a discussion of collection-level and document-level access control models.  Each approach has its advantages and disadvantages---the former approach is simpler while the latter approach makes better use of hardware resources, due to better index compression and higher memory locality.

As it turns out, there are situations where multiple users own documents with overlapping content.  For example, if two users engage in an email conversation and their mail clients quote the previous message when responding to it, then both users will have quoted passages in their corpus---in fact, the fraction of shared passages will increase as a conversation unfolds.  \citet{Broder+2006Indexing} investigated how this phenomenon could be leveraged to reduce the size of a service-wide unified inverted index.  They propose a new data structure called ``document trees'' that captures such sharing of passages of a document and stores but a single instance of the shared passage.  

Respecting users' privacy and preventing any leakage of personal information---either directly or indirectly, say by machine-learned models memorizing individual training examples---is both unethical and may be (depending on the jurisdiction) illegal. There are industry-wide best practices to avoid such leakage.  Here is a non-exhaustive list of such best practices: (1) all personal data must be encrypted both while at rest (i.e., persistently stored) and in flight (i.e., travelling over a network link); (2) each user has a separate encryption key, and user keys are escrowed by a highly secure key management service; (3) no individual has access to the key management services, to prevent insider attacks; (4) only cryptographically signed code can access the key management services; (5) all code is audited before being signed, again to prevent insider attacks; (6) privacy-preserving techniques such as $k$-anonymity (with high $k$) or differential privacy are used to prevent ML models from memorizing individual training examples; and (7) no personal data is retained any longer than required to provide a service to the user.

\section{Beyond Files and Email: Human Digital Memory}
\label{sec:digital-memory}

Taking a step back, the assets contained in a personal collection can be viewed as an externalized extension of that person's memory. The idea of using technology to support such externalized knowledge and to extend a person's memory can be traced back to Vannevar Bush's seminal essay ``As we may think'' \citep{Bush1945AsWeMayThink}, in which he proposed a device termed the {\em Memex} (for ``Memory Extension'') that stores and surfaces all of the information relevant to its user.  Bush wrote this essay in 1945, at the very dawn of the computer age, and so his suggestions for how a Memex could be implemented involved microfilm and mechanical devices, not digital computers. However, the core idea of the essay, that technology can be used to amplify human cognitive ability, is as seductive today as it was in 1945.

Bush's vision provided a source of inspiration to generations of computer scientists, including J. C. R. Licklider and Doug Engelbart.  Licklider wrote the 1960 article ``Man-computer symbiosis'' \citep{Licklider1960Symbiosis} which argued that computers are tools that can amplify human intellect; and as a government administrator he went on to promote research into much of the core technologies of today's information age: computer networking, accessible user interfaces, and the like. His prot{\'e}g{\'e} Doug Engelbart authored a 1962 essay on ``Augmenting human intellect'' \citep{Engelbart1962Augmenting}, founded the SRI Augmentation Research Center (ARC), and created the ``oN-Line System'' (NLS), the blueprint of the modern personal computer that pioneered graphical user interface, the mouse, hypertext, and video conferencing.  

Almost 50 years after Bush articulated the Memex vision, with computers having entered the mainstream of society and psychologists having developed theories of memory, \citet{Lamming+1994Prosthesis} revisited the idea of the Memex. Their paper ``The design of a human memory prosthesis'' synthesizes insights from these two disciplines and argues that the time has come when memory extensions are feasible. The paper describes the PARC Tab, an early example of a pocket-sized computer that could be used for note-taking and that connected wirelessly to a larger computer.  The authors argue that future memory prostheses should be aware of their environment, capture data automatically, allow for active note-taking, and support retrieval with context awareness.

\citet{Bell2001PersonalDigitalStore} laid out a vision for CyberAll, a system intended to digitize and store all of a user's personal information---music, photos, videos, writings, business transactions, and so on. The paper sketches the Windows-based prototype, describes the author's efforts to digitize content (e.g. paper books, vinyl records, and 8mm film), ruminates on the lifetime of physical media and file formats, and argues that continuous advances in storage technology make it economically feasible to store and archive all of a person's information.  CyberAll evolved into MyLifeBits \citep{Gemmell+2006MyLifeBits}, and the prototype system was indeed populated with much of Bell's personal archive.

LifeLogs are an offshoot of human memory prostheses. The basic idea is to record all of a user's activities, including computer-mediated activities such as surfing the Web, but also continuous capture of physical reality, such as recording a person's location and the sounds and views that they experience. \citet{ChenJones2010Augmenting} investigate how such lifelogs advance the original vision of human memory augmentation. 

\section{Future Directions for Personal Information Retrieval}
\label{sec:future-directions}

To conclude, we would like to identify several directions for future research into personal collections.  The first line of research revolves around seamless retrieval from public and personal collections, the second around moving from asset retrieval to task assistance, and the third focused on how personal collections can be integrated into virtual assistants.

With the emergence of cloud-based computing services, personal collections are evermore likely to be stored in the cloud and retrieved through a cloud-based service (with optional caching on users' devices to cope with periods of disconnected operations). In such a setting, a query to a search service can easily surface both public and personal items. The research challenge is to aggregate the results of such a federated search service in a way that is most useful to its users.

People usually perform searches in the pursuit of an overarching goal. Within the Information Retrieval community, there is renewed interest in how retrieval services can go beyond delivering ``ten blue links'' in helping people to achieve these goals.  For example, imagine being able to seamlessly retrieve all of your tax-related documents or to find your friend's address by issuing a single query. Recent advances in natural language processing have demonstrated that it is possible to build such advanced information seeking capabilities on top of general-purpose large language models.  However, integrating personal information into such models in a way that is economically feasible is a wide-open problem.

As virtual assistants such as Google Assistant and Amazon Alexa become more pervasive, it is natural to believe that such assistants will be able to interact more and more with personal collections in the future. Some assistants already have limited access to personal information, for example, to provide access to contacts, calendar information, and so on. However, there is significant opportunity to develop advanced assistive search and intelligence capabilities over personal data that go well beyond what is possible today. This presents many interesting research opportunities at the intersection of personal collections and conversational search.

\backmatter
\bibliography{acm-book}
\chapter*{Authors’ Biographies}

\textbf{Michael Bendersky} is an engineering director at Google DeepMind. He is currently managing a team whose mission is improving algorithms, models, and metrics for information discovery across Google products. Michael holds a Ph.D. from the University of Massachusetts Amherst and is a distinguished member of the ACM. Michael has coauthored over a hundred publications in the areas of information retrieval, natural language processing, and machine learning. He coauthored two books in the \textit{Foundations and Trends in Information Retrieval} series: \textit{Information Retrieval with Verbose Queries} and \textit{Search and Discovery in Personal Email Collections}.

\medskip
\noindent
\textbf{Donald Metzler} is a senior staff research scientist at Google DeepMind. Prior to that, he was a research assistant professor at the University of Southern California and a senior research scientist at Yahoo! He is a distinguished member of the ACM and the recipient of multiple Test of Time and Best Paper awards. He has served as the program chair of the WSDM, ICTIR, and OAIR conferences and sat on the editorial boards of all the major journals in his field. He has published over a hundred research papers, has been awarded 10 patents, and is a coauthor of \textit{Search Engines: Information Retrieval in Practice}. He currently leads a research group focused on a variety of problems related to large language models, machine learning, and information retrieval.

\medskip
\noindent
\textbf{Marc Najork} is a distinguished research scientist at Google DeepMind. Previously, he was a senior director of research engineering at Google Research. Before joining Google in 2014, Marc was a principal researcher at Microsoft Research (2001–2014) and prior to that a researcher at the DEC/Compaq Systems Research Center (1993–2001). Marc earned a Ph.D. in computer science from the University of Illinois. He is an AAAS Fellow, an ACM Fellow, an IEEE Fellow, and a member of the SIGIR Academy. He has coauthored over a hundred research papers and holds 36 US patents. His service activities include editor-in-chief of the \textit{ACM Transactions on the Web} (2011–2014), news board cochair of the \textit{Communications of the ACM} (2008–2014), member of the ACM Publications Board (since 2020), conference chair of WSDM 2008, and program cochair of WWW 2004, WWW 2021, and WSDM 2025.

\medskip
\noindent
\textbf{Xuanhui Wang} is an engineering manager at Google DeepMind. He holds a Ph.D. and an M.S. from the University of Illinois at Urbana-Champaign and a B.E. from the University of Science and Technology of China. His research interests are in information retrieval, machine learning, and natural language processing. For learning-to-rank, he has open sourced two popular deep learning libraries: TFRanking on TensorFlow and Rax on Jax. He has published over seventy research papers in top IR/ML/NLP conferences and won several Best Paper awards: Best Short Paper Nomination of ICTIR (2020), Best Short Paper Award of ICTIR (2019), Best Paper Honorable Mention of WSDM (2017), and Best Paper Award in WSDM (2011). During his career, he has co-invented over 10 US patents.

\end{document}